\documentclass[final,5p,twocolumn]{elsarticle}

\usepackage{graphicx}
\usepackage{tabularx}
\usepackage{dcolumn}
\usepackage{color}
\usepackage{nicefrac}
\usepackage{amsmath}
\usepackage[T1]{fontenc}

\newcommand{\Bhf}{\ensuremath{B_{\mathrm{hf}}} }
\newcommand{\LaBr}{LaBr$_3$ }
\newcommand{\element}[2]{\ensuremath{{}^{#2}\textrm{#1}}}

\journal{Physics Letters B}

\begin{document}

\begin{frontmatter}
\title{Shape polarization in the tin isotopes near $N=60$ from precision $g$-factor measurements on short-lived $11/2^-$ isomers} 
\author[anu,ornl,doe]{T.~J.~Gray}
\fntext[doe]{This manuscript has been authored by UT-Battelle, LLC, under contract DE-AC05-00OR22725 with the US Department of Energy (DOE). The US government retains and the publisher, by accepting the article for publication, acknowledges that the US government retains a nonexclusive, paid-up, irrevocable, worldwide license to publish or reproduce the published form of this manuscript, or allow others to do so, for US government purposes. DOE will provide public access to these results of federally sponsored research in accordance with the DOE Public Access Plan (http://energy.gov/downloads/doe-public-access-plan).}
\affiliation[anu]{Department of Nuclear Physics and Accelerator Applications, Research School of Physics, Australian National University, Canberra, ACT 2601, Australia}
\affiliation[ornl]{Physics Division, Oak Ridge National Laboratory, Oak Ridge, Tennessee 37831, USA}
\author[anu]{A.~E.~Stuchbery}
\author[york,itp]{J.~Dobaczewski}
\affiliation[york]{School of Physics, Engineering and Technology, University of York, York YO10 5DD, United Kingdom}
\affiliation[itp]{Institute of Theoretical Physics, Faculty of Physics, University of Warsaw, PL-02-093 Warsaw, Poland}
\author[koln]{A.~Blazhev}
\affiliation[koln]{
department={Institut f\"ur Kernphysik},
organization={Universit\"at zu K\"oln},
address={Z\"ulpicher Sra{\ss}e 77, D-50937, K\"oln, Germany}
}
\author[anu,tur]{H.~A.~Alshammari}
\affiliation[tur]{College of Science and Arts in Turaif, Physics Department, Northern Border University, Saudi Arabia}
\author[anu]{L.~J.~Bignell}
\author[york,lyon]{J.~Bonnard}
\affiliation{School of Physics, Engineering and Technology, University of York, York YO10 5DD, United Kingdom}
\affiliation[lyon]{
department={Universit{\'e} de Lyon, Institut de Physique des 2 Infinis de Lyon},
organization={IN2P3-CNRS-UCBL},
address={4 rue Enrico Fermi, 69622 Villeurbanne, France}}
\author[anu]{B.~J.~Coombes}
\author[anu]{J.~T.~H.~Dowie}
\author[anu]{M.~S.~M.~Gerathy}
\author[anu]{T.~Kib\'edi}
\author[anu]{G.~J.~Lane}
\author[anu]{B.~P.~McCormick}
\author[anu]{A.~J.~Mitchell}
\author[anu]{C.~Nicholls}
\author[anu]{J.~G.~Pope}
\author[erl]{P.-G.~Reinhard}
\affiliation[erl]{
department={Institut f\"ur Theoretische Physik II},
organization={Universit\"at Erlangen-N\"urnberg},
address={91058 Erlangen, Germany}}
\author[anu]{N.~J.~Spinks}
\author[anu]{Y.~Zhong}

\date{\today}
\begin{abstract}
The $g$ factors of $11/2^-$ isomers in semimagic $^{109}$Sn and $^{111}$Sn (isomeric lifetimes $\tau = 2.9(3)$~ns and $\tau = 14.4(7)$~ns, respectively) were measured by an extension of the Time Differential Perturbed Angular Distribution technique, which uses \LaBr detectors and the hyperfine fields of a gadolinium host to achieve precise measurements in a new regime of short-lived isomers. The results, $g(11/2^-; {^{109}\textrm{Sn}}) = -0.186(8)$ and $g(11/2^-; {^{111}\textrm{Sn}}) = -0.214(4)$, are significantly lower in magnitude than those of the $11/2^-$ isomers in the heavier isotopes and depart from the value expected for a near pure neutron $h_{11/2}$ configuration. Broken-symmetry density functional theory calculations applied to the sequence of $11/2^-$ states reproduce the magnitude and location of this deviation. The $g(11/2^-)$ values are affected by shape core polarization; the odd $0h_{11/2}$ neutron couples to $J^{\pi}=2^+,4^+,6^+...$ configurations in the weakly-deformed effective core, causing a decrease in the $g$-factor magnitudes.
\end{abstract}

\end{frontmatter}

\par
The nucleus is a self-organizing strongly interacting quantum many-body system that displays a variety of behaviors ranging from few-nucleon to collective excitations. In many heavy nuclei, simple patterns in the energy levels are observed despite the underlying complexity of the individual nucleon motion. For nuclei near the magic numbers of the nuclear shell model, 2, 8, 20, 28, 50, 82, and 126, the low-excitation structures are usually associated with the motions of valence nucleons outside the closed shells, which are considered inert. Far from the magic numbers, sequences of excited states associated with the rotations of a deformed spheroid are observed. An active area of research concerns the emergence of collective structures from the apparently few-nucleon excitations of nuclei with proton ($Z$) and/or neutron ($N$) numbers close to the magic numbers. Proton-neutron interactions which induce weak collectivity are essential to explain these transitional systems~\cite{Sb129, Allmond2017}.

\par
The focus here is on the semimagic tin isotopes ($Z=50$), which have a closed proton shell. Pairing correlations between the valence neutrons mean that the low-excitation states of the even-$A$ isotopes are determined by a single broken pair of neutrons, with the remaining nucleons coupled to zero angular momentum, whereas the lowest-excitation structures of the odd-$A$ isotopes are determined by the single-particle states available to the unpaired nucleon; see Ref.~\cite{physics4030048} for a recent review. 
In the odd-$A$ Sn isotopes between $^{100}$Sn and $^{132}$Sn, the lowest 11/2$^-$ state is expected to represent a near pure neutron $0h_{11/2}$ orbit, not only because of the pairing correlations, but also because it has negative parity, whereas the other single-particle orbits in the valence space ($1d_{5/2}$, $0g_{7/2}$, $2s_{1/2}$, and $1d_{3/2}$) have positive parity. However, the uncontested purity of the neutron $0h_{11/2}$ orbit does not preclude the specific polarization effects it may exert on the remaining neutrons and closed-shell protons.

\par
\begin{figure}[h]
  \centering
  \includegraphics[width=\columnwidth]{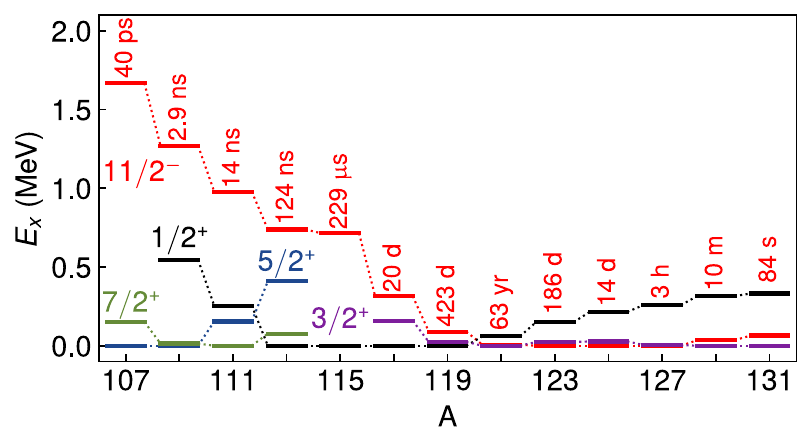}
    \caption{Energies of selected states across the odd-$A$ Sn isotopic chain, showing the sequence of long-lived $11/2^-$ states and their mean lifetimes.}
    \label{fig:energy_syst}
\end{figure}

A sequence of low-lying $11/2^-$ states is observed from $^{109}$Sn to $^{131}$Sn, with meanlives between $\tau = 2.9(3)$~ns and 63.3(7)~years~\cite{Kaubler1995,Rech2002}, as shown in Fig.~\ref{fig:energy_syst}. For the isotopes between $^{113}$Sn and $^{131}$Sn the magnetic moments are consistent with a near pure $\nu h_{11/2}$ configuration; the $g$~factors (magnetic moment divided by angular momentum) are near constant from $^{113-131}$Sn~\cite{Dimmling1974,Brenn1974,Gumprecht1972,Bruer1971,Anselment1986,Blanc2005,Yordanov2020}, with $g \approx -0.24$ (the single-particle $g$ factor with the spin contribution $g_s = 0.7 \times g_{s_{\mathrm{free}}}$). The literature value in $^{111}$Sn is also consistent with this scenario, but the uncertainty is large~\cite{Brenn1974}.

\par
This Letter reports precise measurements of the $g$~factors of the 11/2$^-$ isomers in $^{109}$Sn and $^{111}$Sn, based on new developments in the Time Differential Perturbed Angular Distribution (TDPAD) technique~\cite{Gray2017,Gray2020}: the use of \LaBr detectors, which have excellent timing {\em and} good energy resolution, together with the internal hyperfine fields after in-beam implantation into a ferromagnetic host, opens up a new regime for precise $g$-factor measurements on short-lived excited states ($\tau \approx$ a few~ns). The $g$ factors of the lowest $11/2^-$ states of $^{109}$Sn and $^{111}$Sn with $\tau = 2.9(3)$~ns~\cite{Kaubler1995} and $\tau = 14.4(7)$~ns~\cite{Prade1984}, respectively, were measured relative to $^{113}$Sn with $\tau = 118.5(25)$ ns~\cite{Dimmling1974,Brenn1974}.
A marked deviation from the near-constant $g$~factors of the heavier isotopes was observed. 
Curiously, the deviation occurs at $N \approx 60$, where unexpectedly enhanced $B(E2; 0^+ \rightarrow 2^+)$ strengths were observed in the neighboring even-even Sn isotopes~\cite{Banu2005,Cederkall2007,Vaman2007,Doornenbal2008,Ekstrom2008,Kumar2010,Bader2013,Guastalla2013,Allmond2015,Kumar2017}, which have generally been interpreted as evidence for breaking of the $Z=50$ shell closure in shell model calculations~\cite{Banu2005,Cederkall2007,Vaman2007,Doornenbal2008,Ekstrom2008,Back2013,Doornenbal2014,Togashi2018}.

\par
To gain insight into the origin of the changes in the structures of the $11/2^-$ isomers, we performed broken-symmetry density functional theory (DFT) calculations. This approach has explained magnetic moments of other odd-$A$ isotopes in the region~\cite{(Ver22b)} and it has certain advantages over the shell-model. For example, to produce deformation, shell-model calculations need basis spaces which quickly become prohibitively large, and even large-basis calculations require the use of effective charges and effective $g$-factors to describe the electromagnetic moments. In contrast, the DFT calculations naturally allow deformed nuclear shapes and spin distributions, and use bare electromagnetic operators~\cite{(Sas22b)}. In short, in the DFT, very large single-particle spaces can be readily included and the core polarization self-consistently builds up within many shells below and above the Fermi energy.

\par
The experiments were performed at the Heavy Ion Accelerator Facility at the Australian National University using apparatus described in Ref.~\cite{Stuchbery2019}. Excited states in $^{109,111}$Sn were populated by 58-MeV $^{16}$O induced reactions on $^{96,98}$Mo. Two targets were prepared by evaporating 0.2 mg/cm$^2$ of separated isotope onto annealed gadolinium foils 4 mg/cm$^2$ thick. An additional layer of $^{98}$Mo 0.07~mg/cm$^2$ thick was evaporated onto the $^{96}$Mo target to enable the simultaneous observation of the precessions of the 11/2$^-$ isomers in $^{109}$Sn and $^{111}$Sn. The beams were pulsed into bunches of $\mathrm{FWHM}\approx 1.5$~ns, separated by 107~ns for \element{Sn}{109,111} and by 535~ns for \element{Sn}{113}. The gadolinium foil was polarized by an applied field of $0.1$~T. The direction of this field was reversed periodically. The target was cooled to $\approx 6$~K throughout the experiment.
\par
Five $\gamma$-ray detectors were used: four \LaBr detectors oriented at $\theta_{\gamma} = \pm 45^{\circ}$ and $\theta_{\gamma} = \pm 135^{\circ}$ relative to the beam axis, and a single HPGe detector located at \mbox{$\theta_{\gamma} \approx -90^\circ$}, for monitoring purposes. A PIXIE-16 DGF data acquisition system recorded $\gamma$-ray energies and times from \LaBr and HPGe detectors, as well as times corresponding to the beam pulse~\cite{Tan2008,PIXIE16}. 

\par
The hyperfine field strength was evaluated using the known $g$~factor of the 11/2$^-$ isomer in $^{113}$Sn, populated by a beam of $^{18}$O at 55 MeV on the $^{98}$Mo target. To check that no changes in the hyperfine field occurred during the measurement, the precession frequency was monitored through the sequence of $\approx 2.5$-hr runs over a run time of $\approx 48$~hrs. Moreover, for the $^{111,113}$Sn measurement with the $^{98}$Mo target, the beam was switched twice between $^{16}$O and $^{18}$O. No evidence was found of changes in the observed precession frequencies.
Additional details on the experimental methodology can be found in Refs.~\cite{Gray2017,Gray2020,Gray2021}.

\par
\begin{figure}[h]
  \centering
  \includegraphics[width=\columnwidth]{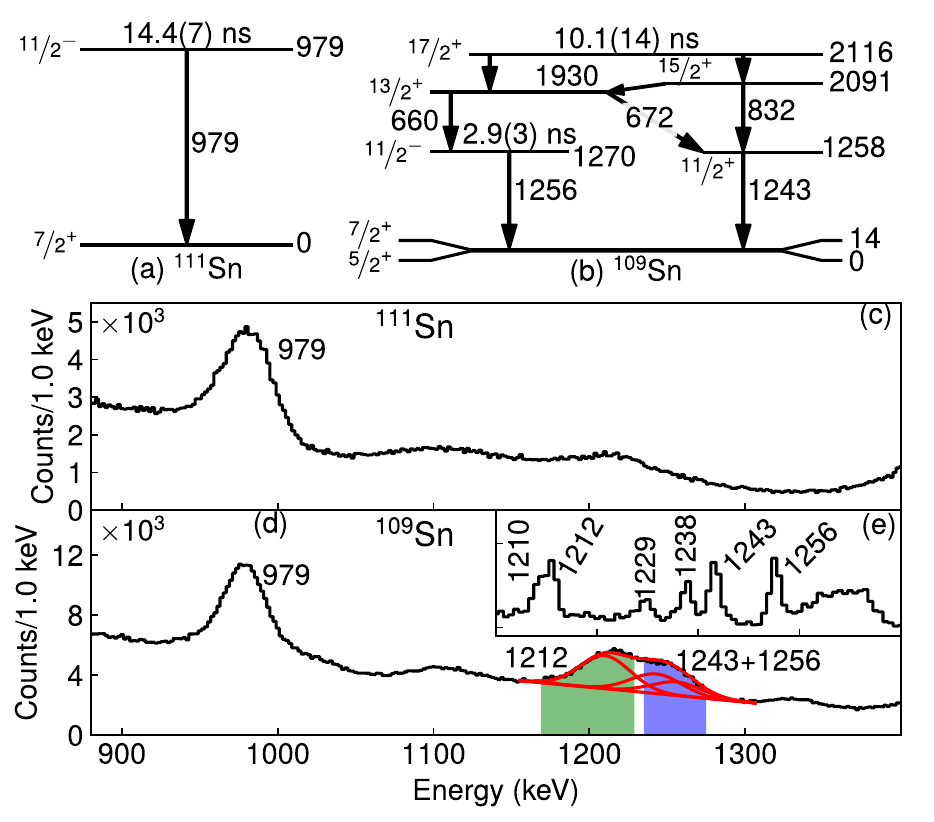}
    \caption{Partial level schemes including isomeric lifetimes of (a) $^{111}$Sn and (b) $^{109}$Sn. Out-of-beam \LaBr energy spectrum following (c) $^{98}\mathrm{Mo}(^{16}\mathrm{O},3n)^{111}\mathrm{Sn}$ and (d)  $^{96,98}\mathrm{Mo}(^{16}\mathrm{O},3n)^{109,111}\mathrm{Sn}$ reactions. The blue and green regions show the energy gates used to construct the 1243-1256 $R(t)$ function, and the 1212 $R(t)$ function, respectively. (e) Region of interest in the HPGe spectrum --- this cannot be gated out of beam due to the poor timing resolution of HPGe detectors. The 1256-keV transition in \element{Sn}{109} is contaminated by the 1243-keV transition in \element{Sn}{109} (see text), and the 1212-keV transition in \element{Sn}{110}. The 1229-keV and 1238-keV ($^{109}$Sn) and 1210-keV ($^{111}$Sn) transitions are all prompt transitions.}
    \label{fig:energy}
\end{figure}
\par
Energy spectra from the \LaBr detectors and relevant parts of the $^{109,111}$Sn level schemes are shown in Fig.~\ref{fig:energy}. The spectra are gated out-of-beam (3 -- 20 ns).

\par
\begin{figure}
  \includegraphics[width=\columnwidth]{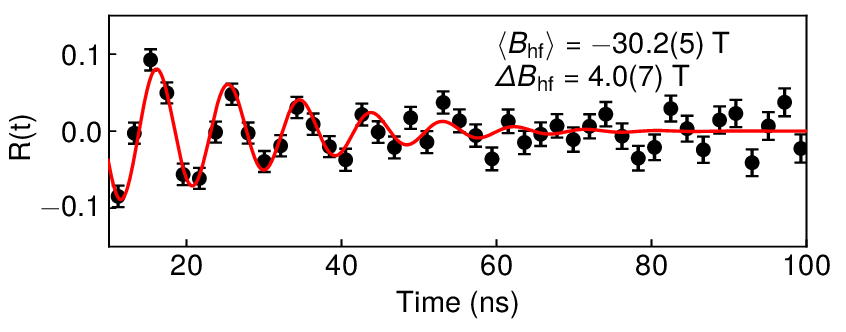}
  \caption{$R(t)$ function for the 661-keV transition depopulating the $11/2^-$ isomer in \element{Sn}{113}. The field distribution was fitted taking $g(\element{Sn}{113}, 11/2^-) = -0.235(2)$~\cite{Brenn1974}.}
  \label{fig:ratio_sn113}
\end{figure}

$R(t)$ functions were constructed in the usual way~\cite{Gray2017,Gray2020}. The oscillations in the $R(t)$ function indicate precession of the excited state. Some damping of the $R(t)$ amplitude was observed: this was attributed to a Gaussian distribution of hyperfine field strengths, which is common for gadolinium hosts~\cite{Gray2017,Gray2020,Raghavan1979}. In both cases, the $R(t)$ functions from the reference isotope (\element{Sn}{113} and \element{Sn}{111}) were used to characterize the mean ($\langle \Bhf \rangle$) and width ($\Delta \Bhf$) of the field-strength distribution. This distribution was then used to fit the $g$ factor(s) of the isotopes of interest. Figure~\ref{fig:ratio_sn113} shows the \element{Sn}{113} $R(t)$ function that was used to fix the \Bhf distribution for the \element{Sn}{111} $g$-factor measurement.
\par
\begin{figure}
  \includegraphics[width=\columnwidth]{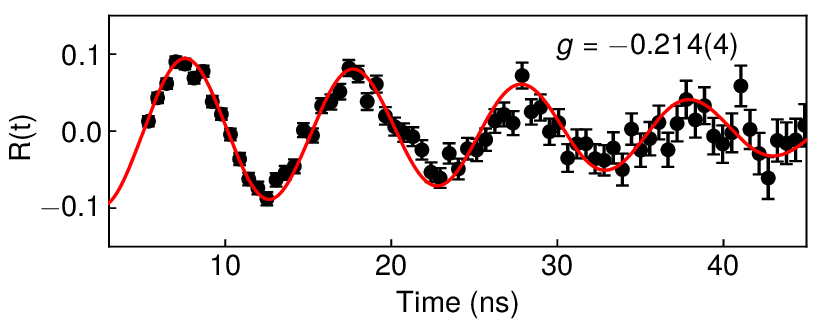}
  \caption{$R(t)$ function for the 979-keV transition depopulating the $11/2^-$ isomer in \element{Sn}{111}. The field distribution was fixed from the $^{113}$Sn measurement.}
  \label{fig:ratio_sn111}
\end{figure}

Figure~\ref{fig:ratio_sn111} shows the $R(t)$ function for the 979-keV transition depopulating the $\tau = 14.4(7)$~ns, $11/2^-$ isomer in \element{Sn}{111}~\cite{Prade1984}. The red curve shows the best fit with $\langle \Bhf \rangle  = -30.2(5)$~T and $\Delta \Bhf = 4.0(7)$~T. This fit gives $g(11/2^-, \element{Sn}{111}) = -0.214(4)$.
\par
\begin{figure}[ht]
  \includegraphics[width=\columnwidth]{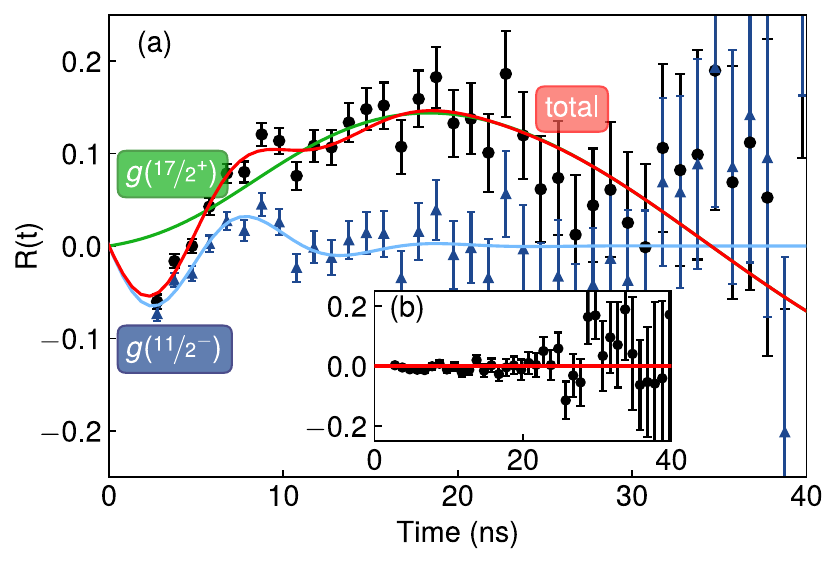}
  \caption{(a) $R(t)$ function for the combined 1243-keV and 1256-keV transitions in \element{Sn}{109}. The solid red line indicates the best fit. The blue and green curves show the oscillations induced in the $R(t)$ function by the $11/2^-$ and $17/2^+$ isomers, respectively. The blue points correspond to the experimental $R(t)$ with the green curve (i.e. the oscillation from the $17/2^+$ state) subtracted. The time zero position is determined precisely from the simultaneous measurement on $^{111}$Sn. (b) $R(t)$ function for the adjacent, 1212-keV region. This peak shows no evidence of precession. }
  \label{fig:ratio_sn109}
\end{figure}

For the $11/2^-$ state in \element{Sn}{109}, the $R(t)$ function is shown in Fig.~\ref{fig:ratio_sn109}(a). In addition to the 1256-keV transition depopulating the $\tau = 2.9(3)$~ns isomer~\cite{Kaubler1995}, contamination from adjacent peaks was present, most notably the 1243-keV transition, which depopulates a $17/2^+$ isomer in \element{Sn}{109} with a lifetime of $\tau = 10.1(14)$~ns~\cite{Kaubler1995} (see Fig.~\ref{fig:energy}). Moreover, the combined 1243-1256-keV peak is only partially resolved from a 1212-keV peak in the LaBr$_3$ detectors. The 1212-keV peak corresponds to the $2^+ \rightarrow 0^+$ transition in \element{Sn}{110} ($\tau = 0.69(6)$~ps). The transition is present out-of-beam because the $2^+$ state is fed by the $6^+$ state at 2478~keV, which has $\tau = 8.1(6)$~ns~\cite{Andrejtscheff1989}. The $R(t)$ function for the 1212-keV transition shows no evidence of precession --- see Fig.~\ref{fig:ratio_sn109}(b) --- and thus does not impact the $R(t)$ analysis for the combined 1243-1256-keV peak.
\par
The effect of the 2116-keV, $17/2^+$ isomer feeding the $11/2^-$ isomeric state through the 660-keV transition was assessed. A limit of $<8\%$ feeding was established from the HPGe spectrum. The effect of this feeding was evaluated using the formalism given by H\"ausser et al. \cite{HAUSSER1976}. The result is to slightly increase the damping of the oscillations, but at a level far below that caused by the distribution of hyperfine fields in the gadolinium host. The shift of the oscillation frequency and thus the effect on the extracted $g$~factor is negligible. Thus, for the purposes of the extracting the $g$~factor from the $R(t)$ function, the decays of the two isomers were considered independent.
\par
The $R(t)$ function for the combination of two independent isomeric states is given by 
\begin{align}
    \label{eq:double_ratio}
    \hspace{-0.5em}R(t) &= \frac{N_0 e^{\nicefrac{-t}{\tau_0}} A_0\sin(2\omega_0 t)
    + N_1 e^{\nicefrac{-t}{\tau_1}}  A_1\sin(2\omega_1 t)}
         {N_0 e^{\nicefrac{-t}{\tau_0}} + N_1 e^{\nicefrac{-t}{\tau_1}}},
\end{align}
where $N_i, \tau_i, A_i$, are the initial population, lifetime, and oscillation amplitude for the states associated with each of the unresolved transitions ($i = 0,1$). The Larmor frequency is $\omega_i = (\mu_N/\hbar) \Bhf g_i$, where $\mu_N$ is the nuclear magneton and $\hbar$ is the reduced Planck constant. The relative populations of the two states were determined from the HPGe spectrum, while the two lifetimes were fixed to the values reported in Refs.~\cite{Kaubler1995,Prade1984}. A time-zero offset, as well as mean hyperfine field-strength $\langle \Bhf \rangle = -31.6(6)$~T and full-width at half-maximum $\Delta \Bhf = 4.6(6)$~T were determined from a fit to the concurrent \element{Sn}{111} $11/2^-$ isomer measurement. Finally, Eq.~\ref{eq:double_ratio} was fitted to the $R(t)$ data with $g_0, g_1, A_0$, and $A_1$ as free parameters. The $R(t)$ function is sensitive to both $g$~factors since the lifetimes of the two states differ by a factor of 3 and the two $g$ factors have opposite signs and magnitudes that differ by a factor of 7. The first part of the $R(t)$ function ($0 - 12$~ns) is most sensitive to $g(11/2^-)$, whereas by the later times ($>12$~ns) $R(t)$ is sensitive only to $g(17/2^+)$.
\par
\begin{figure}[t]
  \includegraphics[width=\columnwidth]{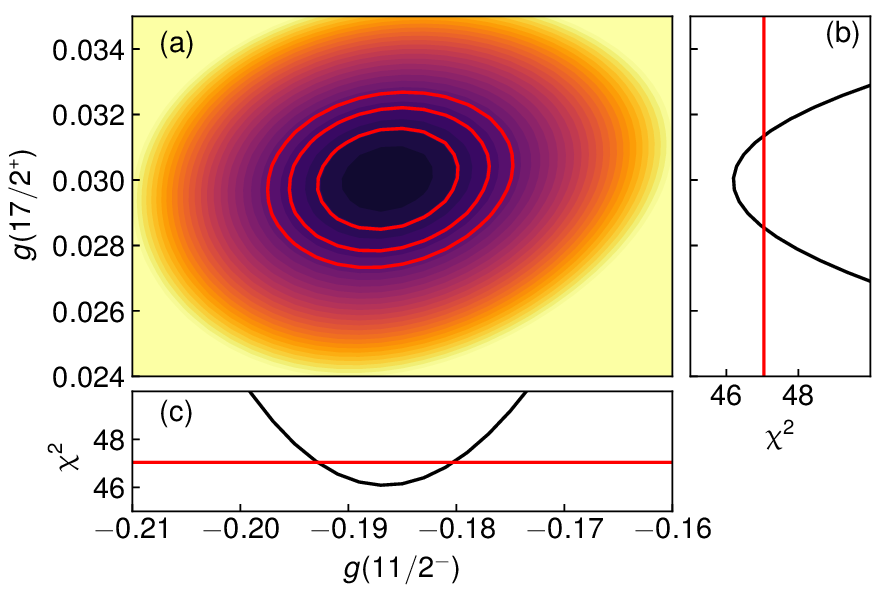}
  \caption{(a) The $\chi^2$ surface for Fig.~\ref{fig:ratio_sn109}. The solid lines correspond to contours at the minimum $\chi^2 + 1, +2$, and $+3$. $\chi^2/\nu = 1.1$ for the best fit. (b) $\chi^2$ curve for $g(17/2^+)$. (c) $\chi^2$ curve for $g(11/2^-)$. Red lines show the minimum $\chi^2 + 1$ level.}
  \label{fig:chi2_sn109}
\end{figure}

The $\chi^2$ surface from the combined fit is shown in Fig.~\ref{fig:chi2_sn109}. Best-fit values are $g(11/2^-) = -0.186(6)$ and $g(17/2^+) = +0.0300(15)$. The uncertainties in the $g$ factors arising from the values that were fixed in this fit were estimated as $2.5\%$ based on a Monte-Carlo simulation; the extracted $g$ factors are not strongly dependent on these values. 
Thus, we obtain $g(11/2^-) = -0.186(8)$ and $g(17/2^+) = +0.030(2)$ for \element{Sn}{109}. The expected configuration of the latter state, \mbox{$\nu [(g_{7/2}^2)_{6^+} \otimes d_{5/2}]_{17/2^+}$} (with the remaining 6 valence neutrons coupled to $J^{\pi} = 0^{+}$), has \mbox{$g = +0.053$}, assuming standard values of \mbox{$g(g_{7/2}) = +0.298$}, and \mbox{$g(d_{5/2}) = -0.536$}, the Schmidt $g$ factors with \mbox{$g_s = 0.7 \times g_{s_\mathrm{free}}$} quenching.
\par
 In addition to the $R(t)$ formed by the peaks at \mbox{$\approx 1250$~keV}, the 832-keV transition can be observed out-of-beam in the \LaBr detectors. The $R(t)$ function from this transition shows an oscillation consistent with the slow component in Fig.~\ref{fig:ratio_sn109} from the $17/2^+$ state, however it is contaminated at short times. Meaningful oscillations were not observed for other peaks; e.g. the 672-keV peak is weak and contaminated by background activity from $\element{Ag}{110} \rightarrow \element{Cd}{110}$ decays [$E(2_1^+; \element{Cd}{110}) = 657$~keV]. 
\par
\begin{figure}[ht]
  \includegraphics[width=\columnwidth]{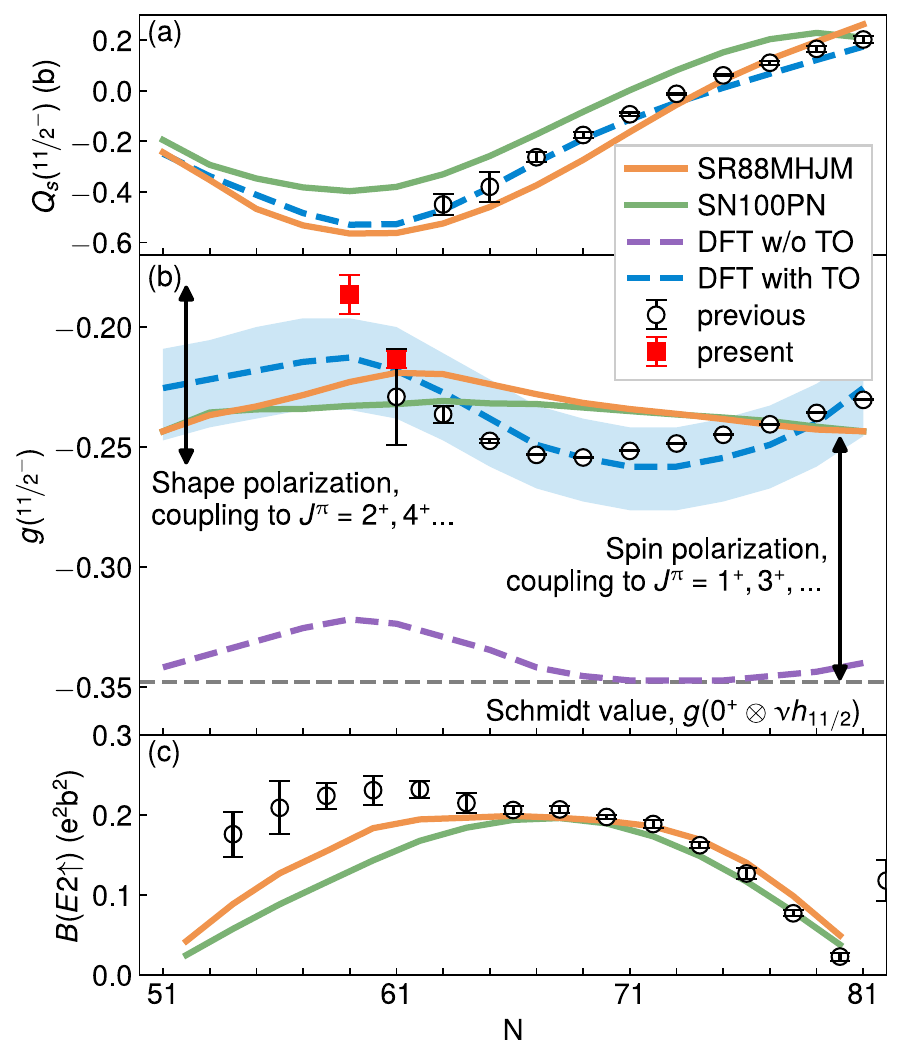}
  \caption{(a) The $11/2^-$ spectroscopic quadrupole moments, data from~\cite{Dimmling1975, Riegel1975, Yordanov2020}. (b) The $11/2^-$ $g$ factors, data from~\protect\cite{Dimmling1974,Brenn1974, Gumprecht1972, Bruer1971,Anselment1986,Blanc2005,Yordanov2020}. Shell-model calculations are plotted in the green and orange solid lines. DFT with the time-odd mean fields included (blue dashed line) or not (purple dashed line) are also shown. The theoretical error band corresponds to the uncertainty in the value of the Landau parameter $g_0'=1.7(4)$~\cite{(Sas22b)}. (c) $B(E2; 0^+ \rightarrow 2^+)$ strengths in the even-even Sn isotopes~\cite{Pritychenko2016}.} 
  \label{fig:systematics}
\end{figure}

The new $g(11/2^-)$ results for \element{Sn}{109} and \element{Sn}{111} are the first indications of a departure from the nearly constant $g(11/2^-) \approx -0.243$ found in heavier Sn isotopes. Figure~\ref{fig:systematics} shows the $g$ factors for \element{Sn}{111-131}~\cite{Dimmling1974,Brenn1974, Gumprecht1972, Bruer1971,Anselment1986,Blanc2005,Yordanov2020}, along with \element{Sn}{109,111} from the present work. Shell-model calculations using the program KSHELL~\cite{Shimizu2019} were carried out using the ``SR88MHJM''~\cite{Faestermann2013,Kavatsyuk2007,Yordanov2018} and ``SN100PN''~\cite{Brown2005} Hamiltonians. Both calculations have closed proton shells, and include $\nu(1d_{5/2}, 0g_{7/2}, 1d_{3/2}, 2s_{1/2}, 0h_{11/2})$ orbitals. The effective spin $g$ factor was quenched to $0.7 \times g_{s_{\mathrm{free}}}$, and effective charges of $e_n = 1.0$ and $0.8$ were used for ``SR88MHJM'' and ``SN100PN'', respectively. 

\par
 Broken-symmetry DFT calculations were performed using the code {\sc{hfodd}} (3.16n)~\cite{(Dob21e),(Dob22)} and the standard Skyrme force UNEDF1~\cite{UNEDF1}. The methodology employed in this work  followed the studies of high-spin isomeric states in heavier elements~\cite{(Bon23a)}. By performing calculations with broken time-reversal symmetry, the time-odd mean fields~\cite{Perlinska2004} generated by the spin-spin two-body force can be included. Its strength was defined by the isovector Landau parameter $g_0'=1.7(4)$~\cite{(Sas22b)}. The time-odd fields lead to a self-consistent evaluation of the spin polarization summed to all orders, in contrast to the first-order effects included in the standard shell-model picture~\cite{Castel1990}.
\par
To obtain the DFT configurations for the $11/2^-$ states, the quasiparticle state having the largest overlap with the neutron Nilsson-like axial orbital $[Nn_z\Lambda{}]K=[505]11/2$ for the angular-momentum projection on the axial-symmetry axis of \mbox{$\Omega=+K=+11/2$} was blocked~\cite{OneQDFT,(Dob09g)}. The angular-momentum symmetry was then restored to the intrinsic HFB states~\cite{(Dob21e),(She21)}, which allowed computation of the spectroscopic magnetic dipole moments $\mu$ and spectroscopic electric quadrupole moments $Q_s$. In the DFT calculations used here, parity symmetry was conserved at the mean-field level and thus negative-parity states of the core did not affect the extracted $Q_s$ or $\mu$ values.

\par
Results of the DFT calculations are shown in Fig.~\ref{fig:systematics}. Those labelled ``UNEDF1 w/o T-odd'' (``UNEDF1 with T-odd''), correspond to the UNEDF1 functional without (with) time-odd mean fields.  The time-odd mean fields quite uniformly reduce the magnitude of the $g$ factors from the single-particle Schmidt value ($g \approx -0.35$) to close to the experimental values, $g \approx -0.243$. This spin polarization is a well-known effect, and is usually included through use of a quenched or effective spin $g$ factor~\cite{Arima1987,Towner1987}. However, the DFT calculations can produce this effect with no need for such quenching. Core polarization contributions from many shells naturally incorporate configurations such as $[g_{9/2}^{-1} \otimes g_{7/2}]_{1^{+}}$~\cite{Arima1987,Towner1987,Brown2005}. 

\par
Both with and without time-odd mean fields, the calculated $g$ factors exhibit a marked dependence on the neutron numbers: this is a result of the odd $0h_{11/2}$ neutron coupling to even-spin configurations of the core~\cite{(Ver22b),(Bon23a)}. The coupling is governed by time-even mean fields induced by the quadrupole-quadrupole interaction. Again, in the DFT, the induced time-even mean fields lead to shape polarization summed up to all orders, in contrast to the so-called second-order perturbative effects included in the shell-model picture~\cite{Castel1990}. The varying coupling to the shape-polarized core is responsible for the decreasing magnitude of the $g$-factor with neutron numbers from $N=69$ to $59$. Notably, this decrease is reproduced in the DFT calculations that do not include time-odd fields; it cannot be attributed to the usual spin-polarization, where spin-orbit partners couple to $1^{+}$, i.e. $[g_{9/2}^{-1} \otimes g_{7/2}]_{1^{+}}$~\cite{Arima1987,Towner1987,Brown2005}. 
The agreement between the DFT calculations and the experimental spectroscopic quadrupole moments is excellent. 

\par{}%
Theoretical predictions are always accompanied by uncertainties related to parameters~\cite{(Dob14)}. The functional UNEDF1 used here generates some dependence of the results on its paring-force parameters. Within the estimated uncertainty of those parameters, the results shown were obtained with pairing-force values increased by 20\% relative to those used in Ref.~\cite{(Bon23a)}, which produced better experimental agreement for both $Q_s$ and $g$. The value of the isovector Landau parameter used in the present calculations ($g_0' = 1.7(4)$~\cite{(Sas22b)}) differs from that used for the In isotopes in Ref.~\cite{(Ver22b)} ($g_0' = 0.82$), which preceded the global analysis of Ref.~\cite{(Sas22b)}. A systematic comparison of the DFT results for $\mu$ and $Q_s$ in the In and Sn isotopic chains is called for. It is noteworthy that the electromagnetic moments have not been used to set the DFT parameters to date.
\par
There has been considerable literature on the increase of $B(E2)$ strength in the Sn isotopic chain near $N=60$~\cite{Banu2005,Cederkall2007,Vaman2007,Doornenbal2008,Ekstrom2008,Kumar2010,Bader2013,Guastalla2013,Allmond2015,Kumar2017,Back2013,Doornenbal2014,Togashi2018}, which is shown in Fig.~\ref{fig:systematics}(c).  The enhancement near $N=60$ has generally been associated with excitations across the $Z=50$ shell gap. While extended-space shell-model calculations can reproduce the $B(E2)$ strengths across the chain (see, e.g., Ref.~\cite{Togashi2018}), they still require effective charges. It is possible that the deviation of the $g(11/2^-)$ values near $N=60$ is related to the observed $E2$ trends, however it will take additional theoretical work to establish the exact relationship between the two phenomena based on models that treat the even-$A$ and odd-$A$ nuclei on an equal footing. Nevertheless, it is evident that the magnetic moments provide a sensitive probe of emerging collectivity in atomic nuclei and provide insights that are complementary to the $E2$ transition strength data.

\par
The present work shows that features of the electromagnetic moments of the semi-magic Sn isotopes require the polarization of shape and spin distributions from single-particle spaces far beyond the valence spaces used in traditional shell-model calculations for the region. The nuclear DFT offers such insights. The unique perspective electromagnetic moments offer on the question of emerging collectivity, beyond what is usually inferred from the $B(E2)$ values in even-even nuclei, has also been demonstrated. In particular, this brings about a question of whether the very notion of the ``core'' should depend on which type of polarizing particle is in action; a comparative analysis of the effective cores in odd indium, antimony, and tin isotopes is very much called for~\cite{(Bon23a)}.

\par
In conclusion, an extension of the TDPAD technique has opened up a new regime of precise $g$-factor measurements for excited states with lifetimes on the order of a few ns. Application of the technique to the $11/2^-$ isomers in $^{109,111}$Sn reveals a remarkable and unexpected deviation from the near-constant $g \approx -0.243$ observed in heavier Sn isotopes. State-of-the-art DFT calculations satisfactorily reproduce the absolute magnitude of the $g$ factors across the entire isotopic chain, including the deviation near $N=60$. The calculations indicate that the $g$-factor variations arise from the core responding to the odd neutron, including excitations well beyond the valence space available to shell-model calculations.

\par{}%
We are also grateful to the academic and technical staff of the Department of Nuclear Physics and the Heavy Ion Accelerator Facility (Australian National University), especially J.~Heighway for assistance with target preparation. We also gratefully acknowledge useful communications from Michael Bender. 
T.J.G, B.J.C, J.T.H.D, M.S.M.G, and B.P.M acknowledge the support of the Australian Government Research Training Program.  This research was supported in part by Australian Research Council grant number DP170701673, as well as the International Technology Centre Pacific (ITC-PAC) under Contract No. FA520919PA138. Support for the ANU Heavy Ion Accelerator Facility operations through the Australian National Collaborative Research Infrastructure Strategy (NCRIS) program is acknowledged. This material is based upon work supported in part by the U.S. Department of Energy, Office of Science, Office of Nuclear Physics under Contract No.~DE-AC05-00OR22725. The publisher acknowledges the US government license to provide public access under the DOE Public Access Plan 
This work was partially supported by the STFC Grant Nos.~ST/P003885/1 and~ST/V001035/1, by the Polish National Science Centre under Contract No.~2018/31/B/ST2/02220, and by a Leverhulme Trust Research Project Grant. We acknowledge the CSC-IT Center for Science Ltd., Finland, for the allocation of computational resources. This project was partly undertaken on the Viking Cluster, which is a high performance computer facility provided by the University of York. We are grateful for computational support from the University of York High Performance Computing service, Viking and the Research Computing team. This research used resources of the Compute and Data Environment for Science (CADES) at the Oak Ridge National Laboratory, which is supported by the Office of Science of the U.S. Department of Energy under Contract No. DE-AC05-00OR22725.

\bibliography{Sn109}

\end{document}